\documentclass[aps,pra,twocolumn,superscriptaddress]{revtex4}
\usepackage{mathrsfs}
\usepackage{epsfig}
\usepackage{graphicx}
\usepackage{amsfonts}
\usepackage[figuresright]{rotating}
\usepackage{amssymb}
\usepackage{amsmath}
\usepackage{dcolumn}
\usepackage{bm}
\usepackage{color}

\def\be{\begin{equation}} \def\ee{\end{equation}}
\def\bea{\begin{eqnarray}} \def\eea{\end{eqnarray}}

\newcommand{\sjtu} {Wilczek Quantum Center and Key Laboratory of Artificial Structures and Quantum
Control, Department of Physics and Astronomy, Shanghai Jiao Tong University, Shanghai 200240, People's Republic of China}

\begin{document}
%\title{Imaginary time crystal: {\clb an exotic phase of quantum matter}}
\title{Persistent oscillations versus thermalization in the quench dynamics of quantum gases with long-range interactions}

\author{Yikai Chen}
\affiliation{\sjtu}

\author{Zi Cai}
%\email{zcai@sjtu.edu.cn}
\affiliation{\sjtu}

\begin{abstract}  % Science journal - 125 word or less; current 119.
Searching for nonthermalized dynamics in interacting quantum systems is not only of fundamental theoretical interest in nonequilibrium quantum physics, but also of immense practical significance  in quantum information processing. In this paper, we study quantum quench dynamics in an  hard-core bosonic model with infinite-range interactions, which have been realized in recent high-finesse cavity experiments.  We show the long-time dynamics of this model can exhibit either undamped oscillations or thermalization depending on the choice of initial states. The long-range nature of the interactions rather than conserved quantities are responsible for such nonergodic dynamical behaviors.
\end{abstract}

%\pacs{05.30.Jp, 75.10.Pq, 02.70.Ss, 03.65. Yz}

\maketitle

%\bibliography{real}
\section{Introduction}
The statistical ensemble description of generic many-particle systems relies on the ergodicity hypothesis, on long timescales, the trajectories of the systems will uniformly cover constant-energy manifold in the phase (Hilbert) space. The macroscopic quantities of physical systems eventually relax to stationary values that are independent of initial conditions other than energy, and can be predicted by statistical ensembles.  Such thermalization phenomena occur if the microscopic Hamiltonian satisfies eigenstate thermalization hypothesis (ETH)\cite{Deutsch1991,Srednicki1994},  which will be stated later on.  However, in special cases such as integrable or many-body localized systems\cite{Basko2006,Oganesyan2007,Znidaric2008,Pal2010,Schreiber2015},  both ergodicity and ETH are broken due to  emergence of many conserved quantities. It is conjectured that such systems might relax to non-thermal states described by  generalized Gibbs ensembles\cite{Rigol2007}.  Recently, Turner {\it et al} proposed an intriguing possibility of the absence of thermalization in a quantum many-body system whose Hamiltonian satisfies ETH, while the ergodicity can be weakly broken due to the presence of special eigenstates (dubbed ``quantum scarred'' states) of the Hamiltonian  instead of conserved quantities\cite{Turner2018}.
 %the suppression of equilibration, (the Fermi-Pasta-Ulam model for instance, see \cite{Berman2005} for a review), or the relaxation to states that strongly depend on the initial state choices.

Ultracold atomic system has provided a highly tunable and well-isolated experimental platform to explore  non-equilibrium quantum many-body physics\cite{Eisert2015}.  A system can be driven out of thermodynamic equilibrium by suddenly quenching\cite{Rigol2008,Polkovnikov2011,Gring2012,Trotzky2012,Alessio2015}, slowly ramping\cite{Braun2015} or periodically driving\cite{Struck2011,Struck2012,Jotzu2014} the parameters in the Hamiltonian.  The typical time scales of ultracold atomic dynamics are much longer than those in other quantum systems ({\it e.g.} solid-state setups), which makes it's much easier to observe and manipulate the non-equilibrium processes in such systems. On the other hand, ultracold atomic systems open up intriguing possibilities to explore the long-range interactions other than Coulomb interactions. Examples include the dipole-dipole interaction between dipolar atoms\cite{Lahaye2009} and molecules\cite{Carr2009},  photon-mediated interaction in high-finesse cavities\cite{Baumann2010, Ritsch2013, Landig2016} and van der  Waals interactions  between  Rydberg  atoms\cite{Heidemann2008,Valado2016}. Due to the breakdown of locality,  long-range interacting quantum systems with nonlocal propagation of information and correlation  usually exhibit non-equilibrium  behaviors that drastically differ from those in their short-range counterparts\cite{Foss2015,Gong2014}.

In this work, we study the quench dynamics of a quantum many-body system with nearest neighboring (NN) hopping but infinite-range interactions, which has been realized in recent high-finesse cavity experiments\cite{Landig2016}. It is shown that the long-time dynamics of this system can exhibit either persistent oscillations or relaxation depending on the choice of the initial states.   Such persistent oscillations are unusual for both generic and integrable quantum many-body systems. In generic cases, scatterings between quasiparticles usually lead to damping for collective oscillations, and make the system equilibrate towards thermal states. For integrable cases, {\it e.g.}, a quadratic fermionic model, the dynamics can be considered as a collective behavior of independent modes with various characteristic frequencies.  A summation over these modes causes a dephasing effect that suppresses oscillations and leads for relaxation\cite{Barthel2008}. In spite of these facts, undamped oscillations in interacting quantum systems has been proposed or observed in plasma physics\cite{Bernstein1957,Gould1967},  ultracold fermions\cite{Yuzbashyan2006,Barankov2006}, and Rydberg atoms\cite{Bernien2017} for different mechanisms. Here we show that the long-range nature of the interactions can gives rise to the undamped oscillations with a characteristic frequency that spontaneously emerges in our model.

\begin{figure*}[htb]
\includegraphics[width=0.32\linewidth,bb=100 61 730 520]{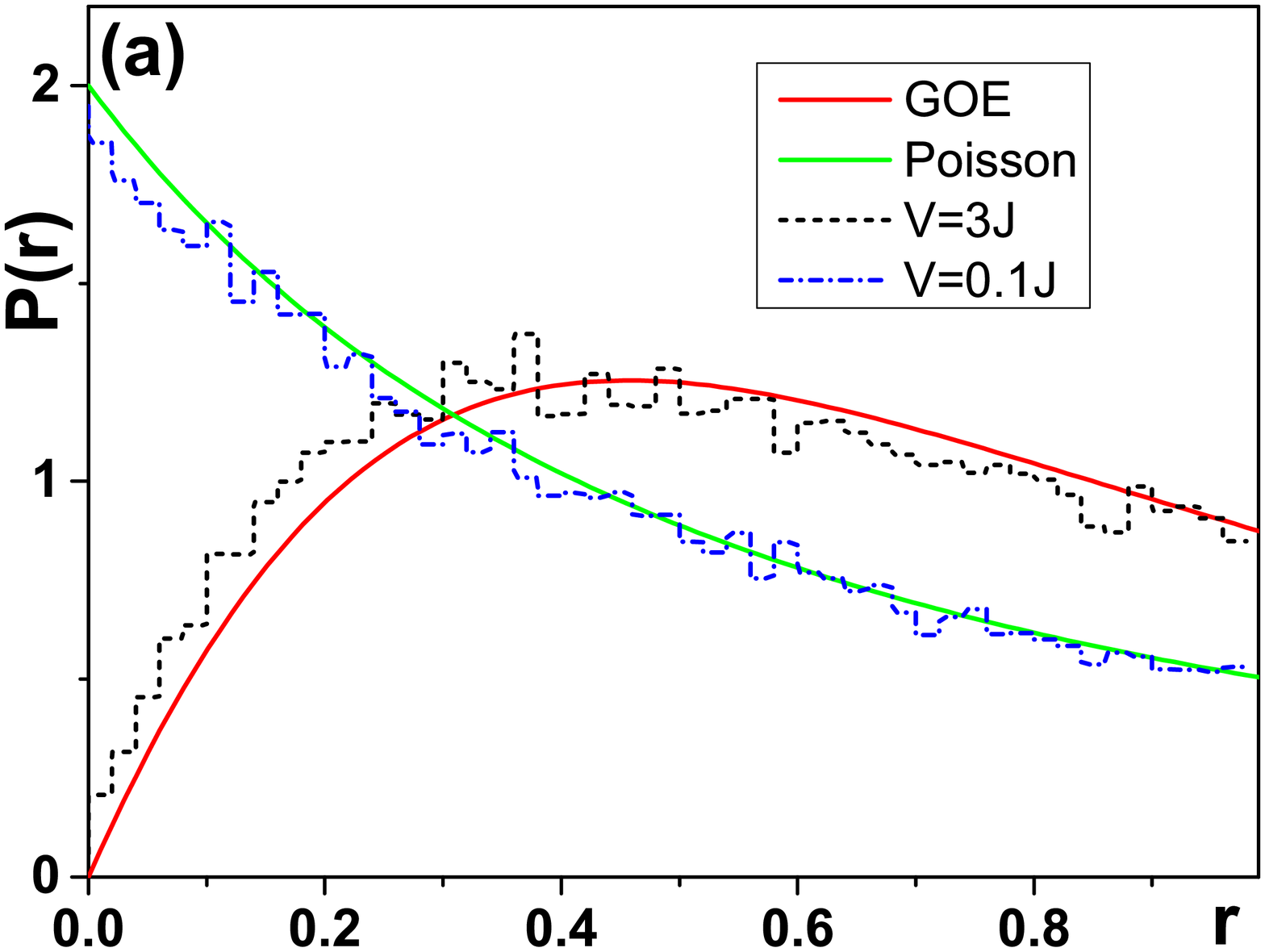}
\includegraphics[width=0.32\linewidth,bb=82 61 704 530]{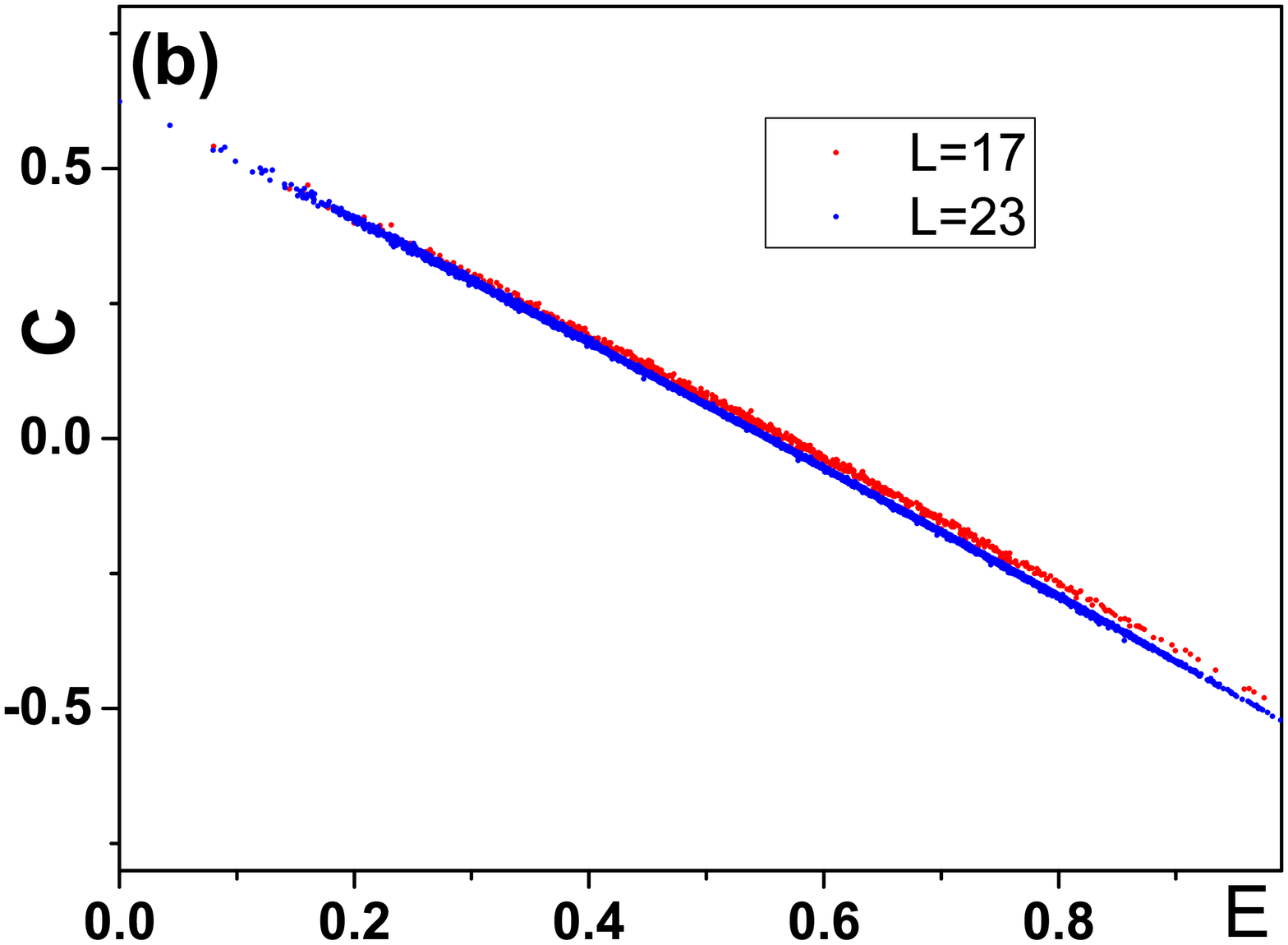}
\includegraphics[width=0.33\linewidth,bb=90 58 733 531]{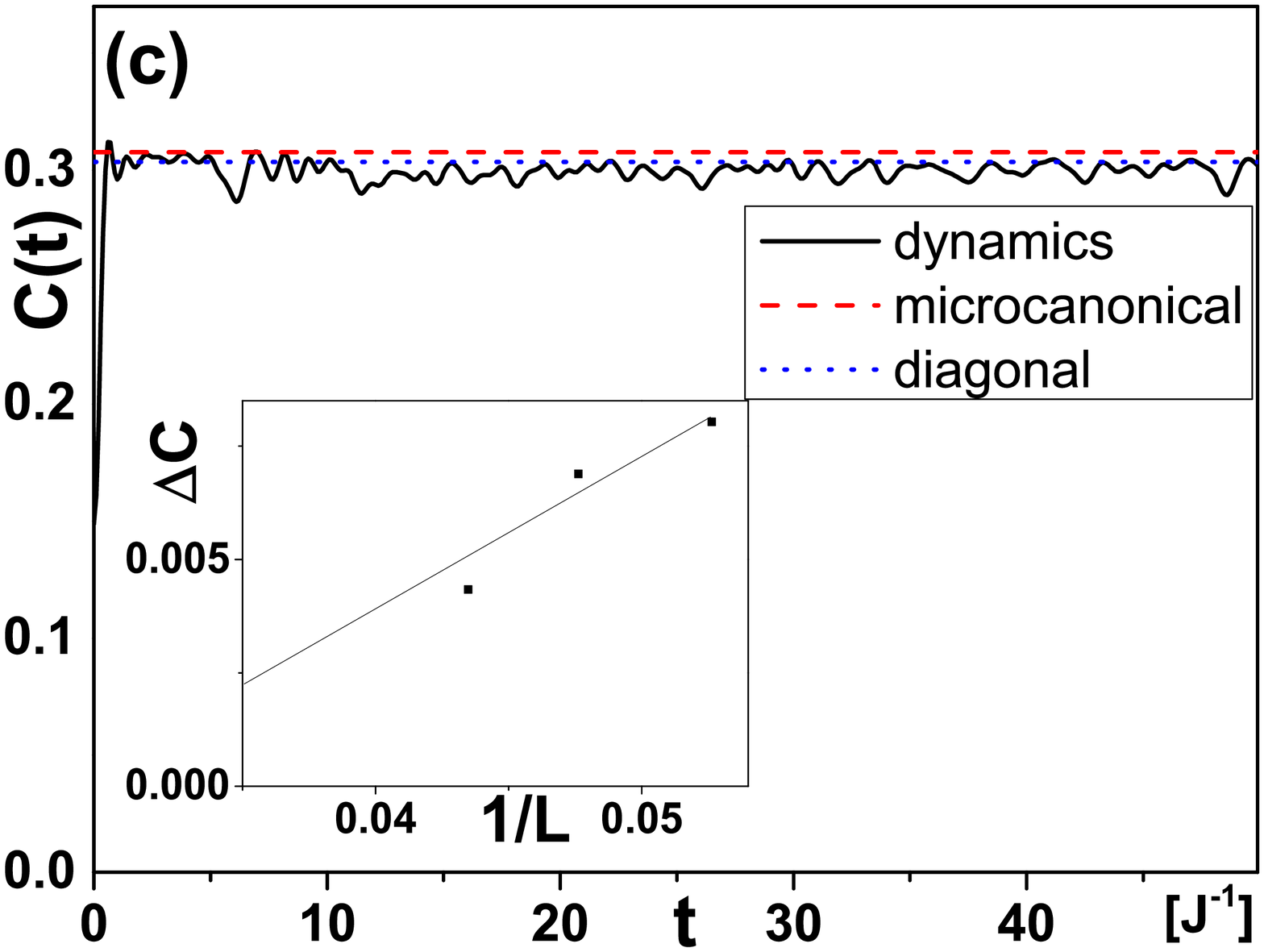}
\caption{(Color online).
(a)Statistics of gaps between the  adjacent energy levels of Hamiltonian.(\ref{eq:Ham}) with $L=23$ and different $V$; (b) The expectation value $C=\langle \Psi_n|\hat{b}_i^\dag\hat{b}_{i+1}|\Psi_n\rangle$  as a function of the normalized eigenstate energy $E=\frac{E_n-E_0}{E_{max}-E_0}$ with the parameter $V=3J$, which resembles a smooth curve; (c) The quench
dynamics of $C(t)=\langle b_i^\dag b_{i+1}\rangle$ with the parameters $V_i=100J$, $V_f=3J$ and $L=23$,  compared with the predictions of the diagonal and microcanonical ensembles. Inset: the finite-size scaling of the difference between the predictions of diagonal and microcanonical ensembles $\Delta C=|C_{diag}-C_{micro}|$.} \label{fig:fig1}
\end{figure*}
\section{Model and method}
The model we consider is an one-dimensional (1D) hard-core bosonic model with infinite-range interactions. The Hamiltonian reads:
\begin{equation}
H=-J\sum_i (\hat{b}_i^\dag \hat{b}_{i+1}+h.c)-\sum_{i<j} V_{ij}(\hat{n}_i-\bar{n})(\hat{n}_j-\bar{n}) \label{eq:Ham}
\end{equation}
where $J$ is the single-particle NN hopping amplitude, $\hat{n}_i=\hat{b}_i^\dag \hat{b}_i$ is the particle number operator at site $i$, $\bar{n}$ is the average particle number per site, which is set to be $\bar{n}=1/2$ (half-filling) throughout the paper. $V_{ij}$ is the long-range interaction between particles on sites $i$ and $j$:  $V_{ij}=\frac VL (-1)^{\parallel i-j\parallel}$ where $V$ is the interaction strength and $L$ is the length of the 1D lattice. The prefactor $\frac 1L$ guarantees the total interacting energy scales linearly with system size. In the following, we consider the periodic boundary condition (PBC), and $\parallel\cdot\parallel$ is defined as the ``shorter distance'' between two sites within a ring:  $\parallel i-j\parallel=|i-j|$ for $|i-j|<L/2$; and $\parallel i-j\parallel=L-|i-j|$ otherwise. Under PBC, the long-range interaction is translational invariant $V_{ij}=V_{\parallel i-j\parallel}$, so is the total Hamiltonian.(\ref{eq:Ham}). The equilibrium properties of similar models have been studied previously\cite{Dogra2016,Sundar2016,Chen2016,Niederle2016,Flottat2017}. At half-filling, the ground state of Hamiltonian.(\ref{eq:Ham}) is always a Mott-insulator with a charge-density-wave (CDW) ordering for any $V>0$. In the following, we prepare the initial state as a ground state of the Hamiltonian with $V=V_i$, then suddenly change the interaction parameter to a different value $V=V_f$ and let the system evolve under this new Hamiltonian. The quench dynamics in such a model has been studied previously based on mean-field method. A non-equilibrium phase transition between different steady states has been discovered\cite{Blass2018,Igloi2018}, while the non-relaxation behavior has not been discussed. For a similar model with infinite-range interactions whose period is incommensurate with the lattice constant, a many-body localized phase has been found in such a translational invariant model\cite{Mondaini2017}.

In the following, we study the long-time dynamics of the model using exact diagonalization (ED) method as well as the self-consistent mean-field analysis, which will be stated later on. To determine the dimension of Hilbert space, we first analyze the Hamiltonian symmetries and  the corresponding conserved quantities, with the help of which we can significantly reduce the matrix dimensions in our ED analysis. The total particle number is conserved, which is fixed to be $\frac L2$ (for even $L$) or $\frac {L-1}2$ (for odd L).  In addition, the system Hamiltonian is invariant under translation and reflection, indicating the conservation of total momentum $K$ and parity $P$ respectively. In our ED calculation, we choose the sectors with $K=0$ and $P=1$. Finally, for the case with even L, there is an additional particle-hole symmetry, and we choose the sector of $\mathcal{P}=1$, with $\mathcal{P}$ the eigenvalue of the operator of particle-hole transformation.  In summary, by choosing a basis that encodes above symmetries, the dimension of the Hilbert space is reduced to $\mathbb{D}=\frac 1{L\times 2\times2}{{L}\choose{L/2}}$ for system with even L, and $\mathbb{D}=\frac1{L\times 2} {{L}\choose{(L-1)/2}} $ for the odd $L$. Thanks to these conserved quantities, we can diagonalize the Hamiltonian with the system size up to $L=24$, with $\mathbb{D}=28968$. In the following, we will focus on two different types of physical quantities: the NN coherence $C(t)=\langle b_i^\dag b_{i+1}\rangle$ and the CDW order parameter $m(t)=\frac 1L \sum_i(-1)^i \langle n_i\rangle$, both of which can be observed experimentally as we will show later on.

 \section{Eigenstate properties}
 Before we turn to the time evolution of our model,  we first study the properties of the eigenstates of the final Hamiltonian with $V=V_f$. We begin with the level statistics of the eigenenergies, which is defined as the ratio of adjacent gaps in the energy spectrum\cite{Oganesyan2007}, $r_\alpha=\min(\delta_{\alpha+1},\delta_{\alpha})/\max(\delta_{\alpha+1},\delta_{\alpha})$, with $\delta_\alpha=E_\alpha-E_{\alpha-1}$ are gaps between adjacent energy levels with ordered eigenenergies $\{E_\alpha\}$. As shown in Fig.\ref{fig:fig1}(a), for a small $V=0.1J$,  the probability distributions of $r_\alpha$  ($P(r)$)  satisfy Poisson distribution, which is due to the fact that it is close to the integrable point of the noninteracting model ($V=0$). However, for an intermediate $V=3J$, $P(r)$ satisfies the Wigner-Dyson distribution\cite{Wigner1955}, which agrees with those in symmetric random matrices belonging to a Gaussian orthogonal ensemble (GOE). One of the most important consequences of  Wigner-Dyson distribution is the energy level repulsion ($P(r)\rightarrow 0$ for small $r$), which is usually considered as a signature of quantum chaotic behavior.

 \begin{figure*}[htb]
\includegraphics[width=0.32\linewidth,bb=100 61 730 520]{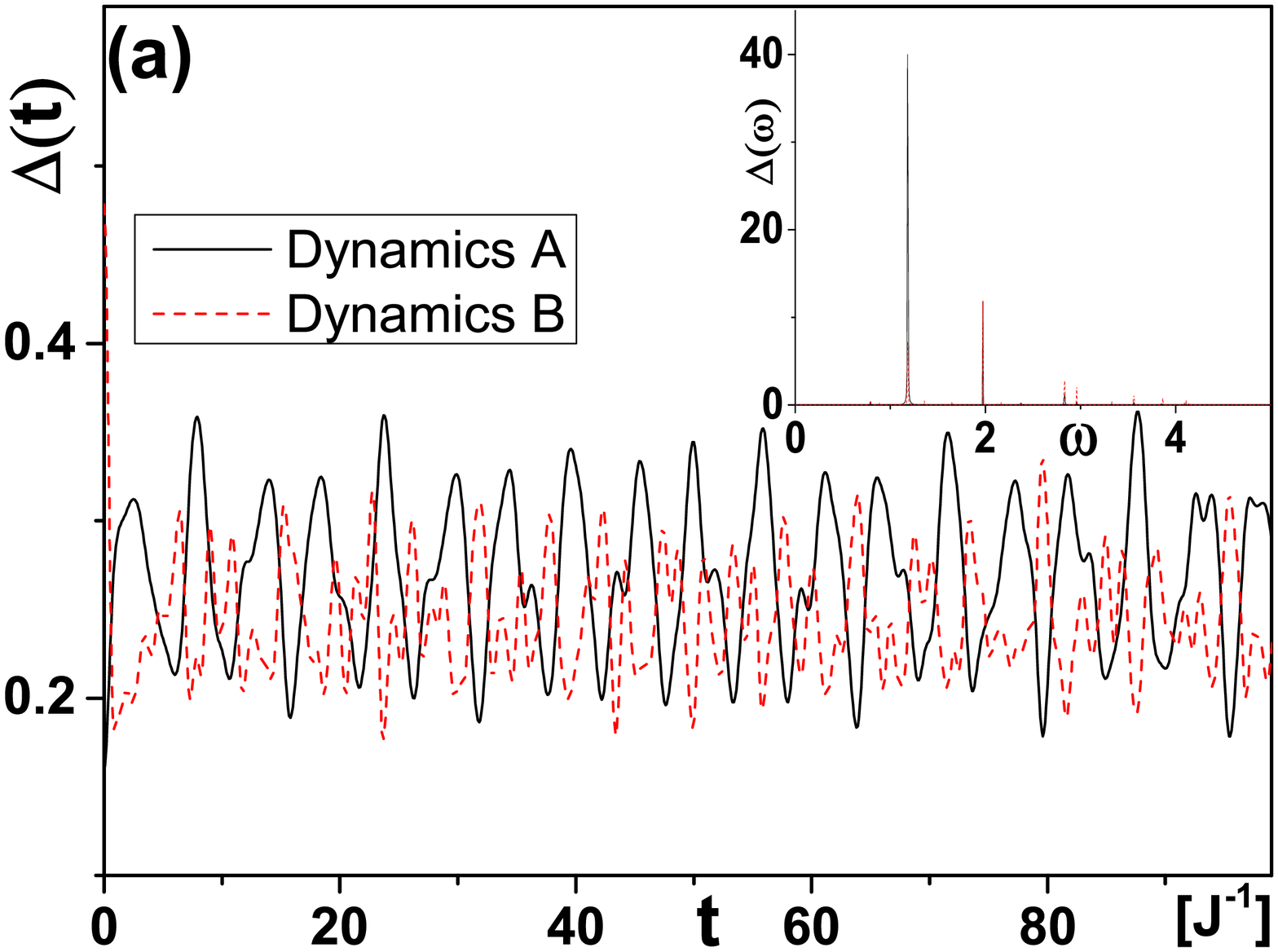}
\includegraphics[width=0.32\linewidth,bb=82 61 704 530]{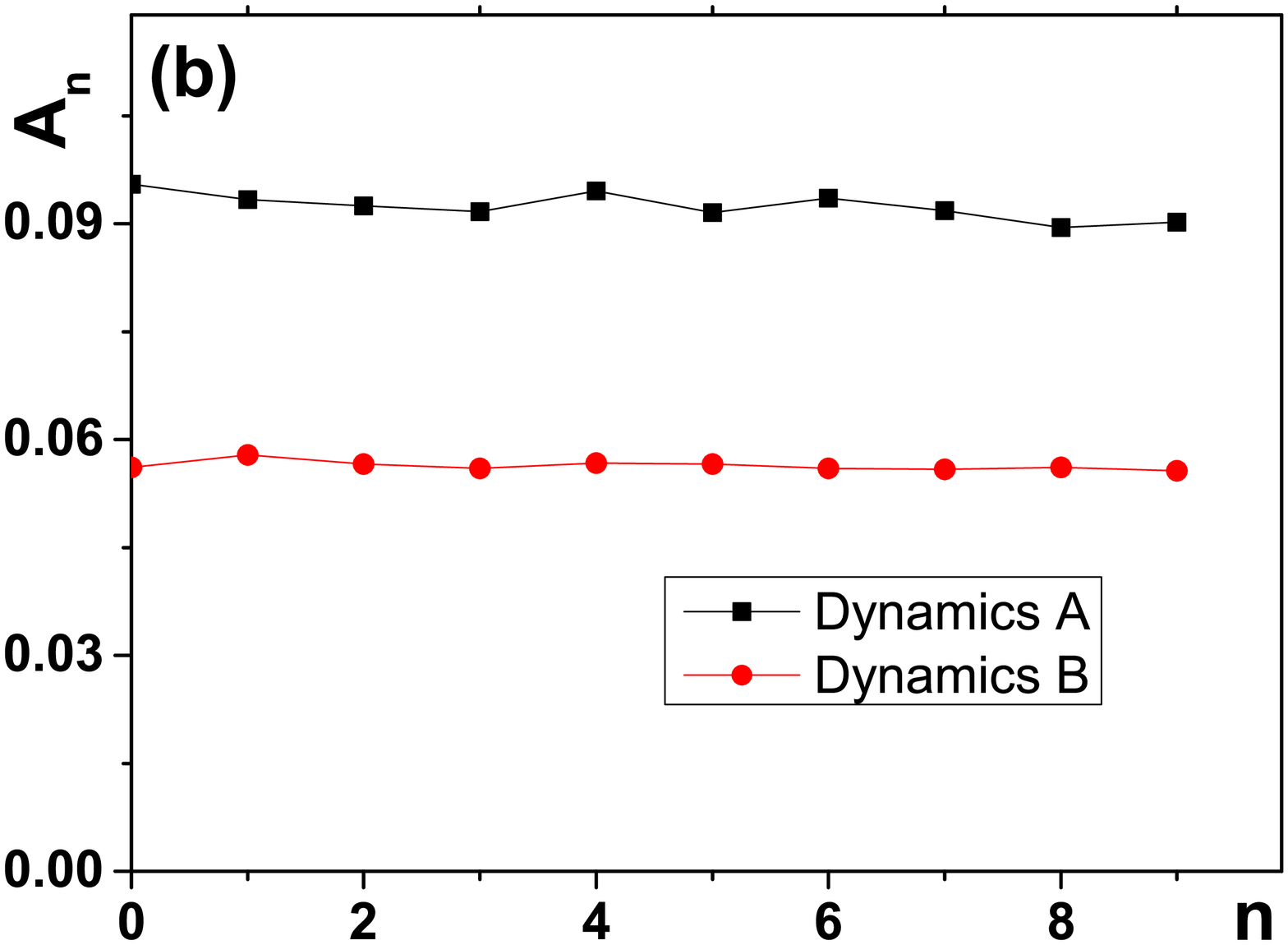}
\includegraphics[width=0.33\linewidth,bb=70 58 713 531]{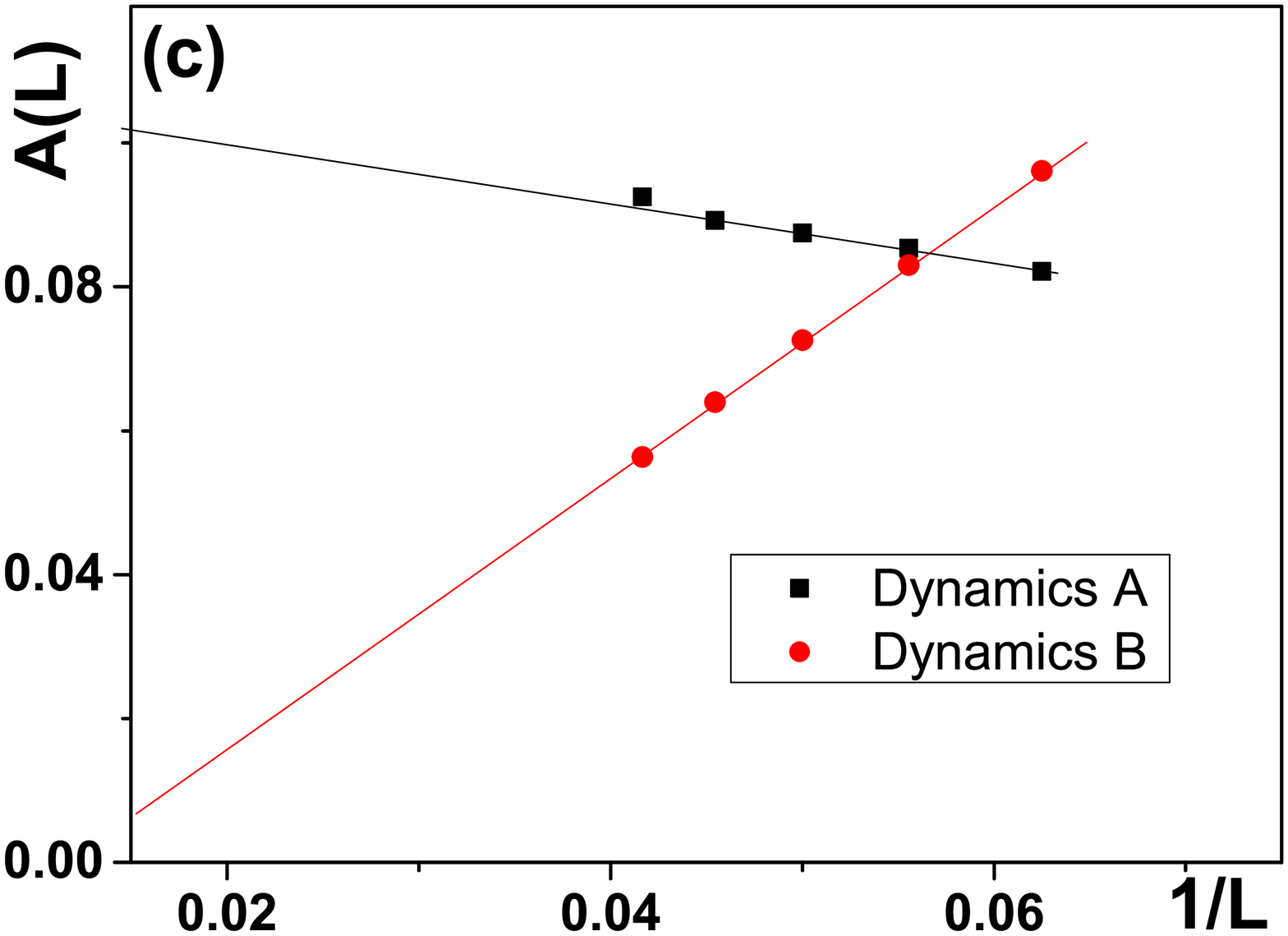}
\caption{(Color online).
(a) Quench dynamics of the CDW order parameter $\Delta(t)$ staring from two different initial states: $V_i=0.75J$ for dynamics A and $V_i=10J$ for dynamics B. Inset: the spectrum functions of the dynamics A and B; (b) The time dependence of the amplitudes of oscillations; (c) The finite size scalings of the oscillation amplitude averaged over the time interval $[0,10^4J^{-1}]$. $L=24$ for (a) and (b), $V_f=3J$ for (a)-(c). } \label{fig:fig2}
\end{figure*}

Next, we check the  eigenstate thermalization hypothesis, which states that for a sufficiently large generic system,  the expectation value of a few-body operator in an eigenstate of the hamiltonian is a smooth function of its eigen-energy, thus equals the microcanonical average of the operator with the same mean energy. We choose the operator as the NN coherence $\hat{C}=\hat{b}_i^\dag \hat{b}_{i+1}$, and its expectation value in different eigenstates as a function of eigenenergies is plotted in Fig.\ref{fig:fig1}(b), from which we can find that for a fixed energy, the variance of $C$ is small and decays with system size, indicating that the ETH is satisfied in our model for sufficiently large systems.  Finally, we check the thermalization of the system. In Fig.\ref{fig:fig1}(c), we plot the time evolution of $C(t)$ starting from the initial state $|\Psi(0)\rangle$  (the ground state of the Hamiltonian with $V_i=100J$) under the new Hamiltonian with $V_f=3J$, and compare the values with those predicted by the diagonal and microcanonical ensembles. The diagonal ensemble value is defined as
\begin{equation}
C_{diag}=\sum_\alpha |c_\alpha|^2\langle \Psi_\alpha|\hat{C}|\Psi_\alpha\rangle
\end{equation}
with $|\Psi_\alpha\rangle$ the $\alpha$-th eigenstate of the final Hamiltonian $H_f$   and $c_\alpha=\langle \Psi_\alpha|\Psi(0)\rangle$. The summation is over all the eigenstates. The microcanonical ensemble value is defined as:
\begin{equation}
C_{micro}=\frac 1{N_\delta}\sum^\alpha_{|E_\alpha-E_0|<\delta}\langle\Psi_\alpha|\hat{C}|\Psi_\alpha\rangle
\end{equation}
where $E_0$ is the average energy, which is conserved during the time evolution. The summation is over the eigenstates with energies in the windows $[E_0-\delta,E_0+\delta]$ (we fix $\delta=0.1J$ throughout this paper), and $N_\delta$ is the total number of such eigenstates. From Fig.\ref{fig:fig1}(c), we can find that these values are close to each other, and their difference decreases with the system size (inset of Fig.\ref{fig:fig1}(c)), which indicates that the system finally thermalizes.

\section{Dynamics of the order parameter}
Now we focus on the dynamics of the CDW order parameter. For a finite system, there is no spontaneous symmetry breaking, thus we use the quantity $\Delta=\sqrt{\frac 1{L^2}\langle[\sum_i (-1)^i\hat{n}_i]^2\rangle}$ to characterize the strength of the CDW order. In the thermodynamical limit,
$\Delta$ extrapolates the CDW order parameter $m=\frac 1L \sum_i(-1)^i \langle n_i\rangle$\cite{Sandvik2010}.  Since we are interested in the initial state dependence of the dynamics,  the finial state Hamiltonian is fixed as $V_f=3J$, and we choose different initial states as the ground states of the Hamiltonian with various $V_i$  ($V_i\neq V_f$). We calculate the time evolution of $\Delta(t)$ for the system with different $L$ starting from two initial states of $V_i=0.75J$ and $V_i=10J$, which are denoted as dynamics A and B respectively throughout this paper.  As shown in Fig.\ref{fig:fig2} (a), for a finite system ($L=24$), both dynamics A and B exhibit persistent oscillations,  while the oscillation amplitude for dynamics A is larger than that of dynamics B.  The spectral functions  $\Delta(\omega)=\frac 1T \int_0^T dt\cos(\omega t) \Delta(t)$ have been plotted the insets of Fig. 2(a), from which we find that $\Delta(\omega)$ for dynamics A and B exhibit sharp peaks at the same positions but with different heights. The positions of the peaks are determined by the energy spectrum of $H_f$, while their heights depends on the initial states.   To characterize the strength of oscillations, we introduce a quantity $A=\bar{\Delta}_{max}-\bar{\Delta}_{min}$, where $\bar{\Delta}_{max}$ ($\bar{\Delta}_{min}$) is the average value of the local maximum (minimum) of $\Delta(t)$ during the time interval of measurement.  To prove the persistence of the oscillation, we further divide the total evolution time interval into N bins, each of which is characterized by $A_n$, the average oscillation amplitude in the time interval $[(n-1)10^3J^{-1},n10^3J^{-1}]$ with $n=1,\cdots,N$. As shown in Fig.\ref{fig:fig2} (b), there are no obvious damping for the oscillations in both dynamics A and B.

Even though for a finite system both dynamics A and B exhibit undamped oscillations, they have different origins: in dynamics B, it is due to the finite size effect, while the persistent oscillation in dynamics A  is intrinsic and can survive in the thermodynamic limit.  To separate the intrinsic persistent oscillation from that induced by the finite size effect, one need perform the finite size scaling. More specifically, we calculate quench dynamics of the system from $t=0$ to $t=T$ with $T=10^4J^{-1}$, and check the finite size dependence of $A$, which is defined in the time interval $[0,T]$. As shown in Fig.\ref{fig:fig2}(c), the oscillation amplitude A slowly increases with the system size $L$ for dynamics A, indicating that such oscillations  can persist even in the thermodynamic limit. On the contrary, the oscillation amplitude decays with system size as $A\sim \frac 1L$ in dynamics B, which indicates that for a sufficiently system size, the system will finally relax to some steady states.

 \begin{figure}[htb]
\includegraphics[width=0.99\linewidth,bb=61 115 770 850]{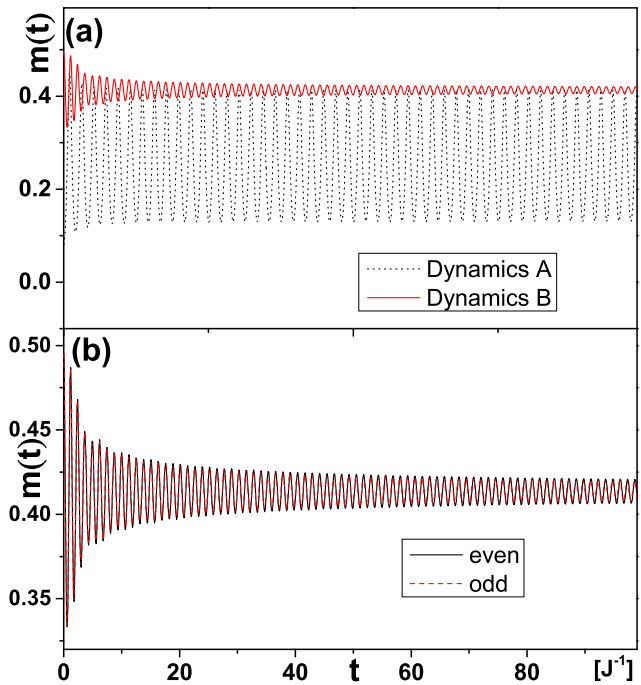}
\caption{(Color online).
(a) Quench dynamics of $M(t)$ calculated by the self-consistent mean-field method with the parameters $V_f=3J$, $L=10^4$ and $V_i=0.75J$ for dynamics A and $V_i=10J$ for dyamics B; (b)A comparison of the quench dynamics of systems with even $(L=500)$ and odd $(L=501)$ lattice sites with parameters  $V_f=3J$ and $V_i=0.75J$.} \label{fig:fig3}
\end{figure}

To further verify the existence of the persistent oscillations in the thermodynamical limit, we perform a dynamical self-consistent mean-field method to study the time evolution of our system.  Under the mean-field approximation, the Hamiltonian with infinite-range interactions is reduced to an 1D hard-core bosonic model with a staggered chemical potential which is self-consistently determined during the time evolution:
\begin{equation}
H_{MF}(t)=-J\sum_i [b_i^\dag b_{i+1}+h.c]+ m(t)V_f\sum_i (-1)^i \hat{n}_i \label{eq:MF}
\end{equation}
where $m(t)=\langle\Psi(t)|\frac 1L(-1)^i\hat{n_i}|\Psi(t)\rangle$ and $|\Psi(t)\rangle$ is the wavefunction of the system at time $t$. The time evolution under the Hamiltonian.(\ref{eq:MF}) can be solved exactly by performing the Jordan-Wigner transformation to transfer 1D hard-core bosons into spinless fermions, and the Hamiltonian.(\ref{eq:MF})  becomes a non-interacting fermionic model. Two different long-time behaviors discussed above can also been found in the dynamics of this mean-field model. As shown in Fig.\ref{fig:fig3} (a), we find a periodic oscillation in dynamics A, while for dynamics B, the amplitude of the oscillation decays with time, and finally converges to a constant value, which agrees with the ED results. In addition, for a small system, there is an even-odd effect: for a system with odd lattice sites, there is  at least one kink in the CDW phase, and the propagation of the kink may make the dynamics of  odd-L system different from that of even-L system.  One may wonder whether such even-odd effect also plays a role in the thermodynamic limit. To clarify this point, we compare the dynamics of the systems with even and odd lattice sites for a sufficiently large system under the mean-field approximation. As shown in Fig.\ref{fig:fig3}(b),  the difference between them are neglectable. The role of the even-odd effect is more or less similar with the boundary condition, which doesn't affect the bulk dynamics for a sufficiently large system.

\section{Experimental realization and detection}
The model we proposed could be experimentally realized by loading bosonic atoms (e.g. $^{87}$Rb) into a deep quasi-1D optical lattice. The strongly onsite repulsive interactions eliminate the possibility of multi-occupancy, thus give rise to the hardcore nature of the bosons. The infinite-range interactions have been realized by coupling the  bosons to a vacuum mode of the cavity via their density operators\cite{Landig2016}. For a large cavity decay rate, the cavity mode can be adiabatically eliminated, which gives rise to the interactions as shown in Hamiltonian.(\ref{eq:Ham}).  The strength of the interaction is controllable by tuning the resonance of the cavity.  For the detections, the CDW order parameter $m(t)$ can be mapped to an intra-cavity photon number, which can be measured via a heterodyne detection\cite{Landig2016}. It can also been directly measured  by a superlattice band-mapping technique\cite{Trotzky2012}. The NN coherence $C(t)$ can be extracted from momentum distributions obtained by the time-of-flight technique\cite{Bouganne2019}. Finally, we estimate the typical time scales of the quench dynamics. For a deep optical lattice with the NN hopping amplitude $J\approx 400$Hz and $V=3J$, the period of persistent oscillations predicted in our model is  roughly $10$ms, a time scale that can be measured in current cold atomic experiments.

Last, we discuss some subtle differences between the experiments and our theoretical model.  In real experimental setups, oscillations are  inevitably damped, because the total energy of the system is not conserved during the evolution because of the dissipations, {\it e.g.} the light leaking out of the cavity. Thus one of the major challenge for the experimental observation is to distinguish the extrinsic damping mechanisms ({\it e.g.} dissipation) from the intrinsic ones ({\it e.g}. thermalization), which is only possible when the damping rates of these two kinds of mechanisms are well separated. Another difference between the experiment and our proposal is that our model is in 1D lattice, while the experimental optical lattice are 2D. A generalization of our model to 2D is straightforward, while our methods are no longer applicable because the hard-core bosons are not equivalent to spinless fermions in a 2D lattice, while the ED method is limited to very small systems. Whether such persistent oscillations can exist in 2D bosonic systems is still an open question and deserve further studies.

\section{Conclusion and outlook}
In summary, we study the quench dynamics of a quantum many-body system with long-range interactions, whose long-time behaviors depend on the initial states. Several problems are left for further studies, including the quantum quench dynamics in 2D systems, which is closely related with current experiment setups\cite{Landig2016}; or in systems with long-range power-law interactions that relevant with experiments of Rydberg atoms and trapped ions. In addition, information propagations in such systems with infinite-range interactions might be fundamentally different from  their short-range counterparts and beyond the restriction of Lieb-Robinson bound due to the absence of locality in the Hamiltonian.

\section{Acknowledgement}
We appreciate insightful discussions with R. Mondaini.  This work is supported  by the National Key Research and Development Program of China (Grant No. 2016YFA0302001), NSFC of  China (Grant No. 11674221 and No.11745006), Shanghai Rising-Star Program, Eastern Scholar Professor of Distinguished Appointment Program and Project of Thousand Youth Talents.

%\bibliography{real}

\end{document}